\documentclass[conference]{IEEEtran}
\IEEEoverridecommandlockouts

\usepackage{cite}
\usepackage{amsmath,amssymb,amsfonts}
\usepackage{algorithmic}
\usepackage{graphicx}
\usepackage{textcomp}
\usepackage{xcolor}
\usepackage{booktabs}
\usepackage{times}

\def\BibTeX{{\rm B\kern-.05em{\sc i\kern-.025em b}\kern-.08em
    T\kern-.1667em\lower.7ex\hbox{E}\kern-.125emX}}
\begin{document}

\title{FAST: Fast Audio Spectrogram Transformer \thanks{© 2025 IEEE.  Personal use of this material is permitted. Permission from IEEE must be obtained for all other uses, in any current or future media, including reprinting/republishing this material for advertising or promotional purposes, creating new collective works, for resale or redistribution to servers or lists, or reuse of any copyrighted component of this work in other works.}}

\author{\IEEEauthorblockN{Anugunj Naman}
\IEEEauthorblockA{\textit{Department of Computer Science} \\
\textit{Purdue University}\\
West Lafayette, Indiana, USA \\
anaman@purdue.edu}
\and
\IEEEauthorblockN{Gaibo Zhang}
\IEEEauthorblockA{\textit{Department of Computer Science} \\
\textit{Purdue University}\\
West Lafayette, Indiana, USA \\
zhan5117@purdue.edu}
}

\maketitle

\begin{abstract}
In audio classification, developing efficient and robust models is critical for real-time applications. Inspired by the design principles of MobileViT, we present FAST (Fast Audio Spectrogram Transformer), a new architecture that combines convolutional neural networks (CNNs) and transformers to capitalize on the strengths of both. FAST integrates the local feature extraction efficiencies of CNNs with the global context modeling capabilities of transformers, resulting in a model that is powerful yet lightweight, well-suited to a real-time or mobile use case. Additionally, we incorporate Lipschitz continuous attention mechanisms to improve training stability and accelerate convergence. We evaluate FAST on the ADIMA dataset, a multilingual corpus towards real-time profanity and abuse detection, as well as on the more traditional AudioSet. Our results show that FAST achieves state-of-the-art performance on both the ADIMA and AudioSet classification tasks and in some cases surpasses existing benchmarks while using up to 150x fewer parameters. 
\end{abstract}


\section{Introduction}
\label{sec:intro}

The landscape of audio classification has undergone significant evolution, primarily driven by the adoption of deep learning models that transform raw audio signals into actionable insights~\cite{gong2021psla,naman,fmaml}. Traditionally, convolutional neural networks (CNNs) have been at the forefront of this transformation, as their inherent spatial locality and translation equivariance have been effective on audio data represented as spectrogram images~\cite{audiocnn,schuller13_interspeech}. In parallel, the success of  architectures in the domain of natural language processing has paved the way for their integration into audio classification tasks, particularly through hybrid models that combine CNNs with self-attention mechanisms to enhance the models’ ability to capture long-range dependencies~\cite{ast,attn_pool,conformer,flashattention,dao2023flashattention2}.

\begin{figure*}
\begin{center}
\includegraphics[width=\textwidth]{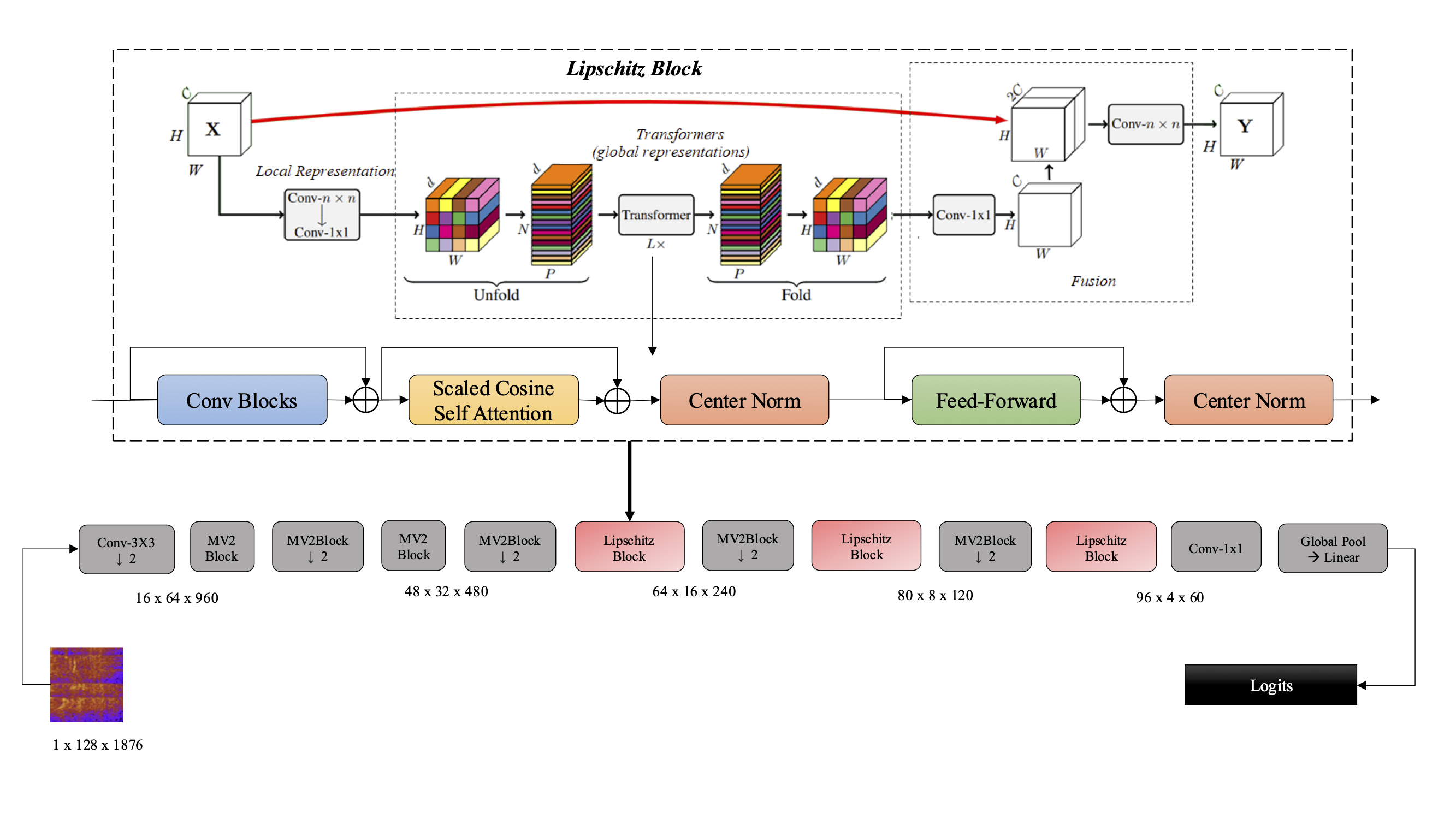}    
\caption{Architecture of the FAST model, highlighting its combination of CNNs and transformers with Lipschitz continuous attention mechanisms. The upper section illustrates the integration of these components, while the lower section presents the full architecture, including MobileNetV2 blocks and Lipschitz-modified blocks for enhanced efficiency and stability.}
\label{fig:fast}
\end{center}
\end{figure*}

However, despite the advancements, the field continues to grapple with the opposing challenges of achieving high performance and optimizing model efficiency—the latter critical for applications requiring real-time processing on resourceconstrained devices. Inspired by recent innovations in visual computing such as the Vision Transformer (ViT)~\cite{vit} and MobileViT~\cite{mobilevit}, which address similar challenges in image processing, we propose a novel architecture: the Fast Audio Spectrogram Transformer (FAST). FAST harnesses the strengths of both CNNs and transformers, integrating them into a unified framework optimized for audio signals.

In creating FAST, we draw inspiration from the MobileViT architecture, which combines the spatial inductive biases of CNNs with the global information processing capabilities of transformers. This synergy allows for a lightweight and efficient model. Moreover, to address the instability issues often encountered during the training of transformer models, we integrate principles from Lipschitz continuity to ensure robustness and stability through the entire training process~\cite{lipsformer}.

This paper introduces FAST, detailing its architecture and the rationale behind combining CNNs with a backbone. We also discuss the integration of Lipschitz continuous components to enhance training stability. The effectiveness of FAST is demonstrated through evaluations on the ADIMA~\cite{adima} and AudioSet~\cite{audioset} datasets which showcase its capabilities in real-world application scenarios for multilingual abusive speech detection and sound event detection, respectively. FAST sets a new benchmark for performance and efficiency on these tasks while using up to 150x fewer parameters, suggesting potential for its efficiencies to push the boundaries of what is possible in broader audio classification.

\section{Related Work}
\label{sec:related}

Recent advancements in audio classification have increasingly explored transformer-based architectures to enhance both performance and efficiency. One of the pioneering models, the Audio Spectrogram Transformer (AST), was the first to rely solely on self-attention mechanisms without convolutional layers, demonstrating that transformers could achieve state-of-the-art results on benchmarks like AudioSet and Speech Commands V2, challenging the dominance of CNNs~\cite{ast}. Following this, HTS-AT introduced a hierarchical token-semantic structure, significantly reducing model size and training time while maintaining high accuracy, particularly in audio event detection, thus outperforming earlier CNN-based models in both efficiency and localization~\cite{htsat}. Another key model in this area is Wav2Vec, which leverages unsupervised pretraining on speech data~\cite{schneider2019wav2vecunsupervisedpretrainingspeech,baevski2020wav2vec20frameworkselfsupervised,conneau2020unsupervised}. Wav2Vec models have been used to extract features from raw audio for various downstream audio tasks, including abuse detection in multilingual corpora like ADIMA~\cite{adima}.  Building on these advances, our proposed FAST architecture integrates the spatial inductive biases of CNNs with the global contextual capabilities of transformers while incorporating Lipschitz continuous attention mechanisms for enhanced training stability. FAST aims to strike a balance between performance and computational efficiency, particularly in resource-constrained environments, and in the following sections, we define the architecture and compare its effectiveness against the AST, HTS-AT, and Wav2Vec-based XLSR-53~\cite{conneau2020unsupervised}.

\section{FAST Architecture}
\label{sec:fast}

FAST employs an efficient convolutional feature extraction process inspired by lightweight architectures like MobileViT. For a given input audio spectrogram $\mathbf{X} \in \mathbb{R}^{H \times W \times C}$, the model applies a $3 \times 3$ convolution followed by a point-wise ($1 \times 1$) convolution, resulting in a transformed tensor $\mathbf{X}_L \in \mathbb{R}^{H \times W \times d}$, where $d > C$. The $3 \times 3$ convolution encodes local spatial features, while the point-wise convolution projects the input into a higher-dimensional space by learning linear combinations of the channels:
\begin{equation}
    \mathbf{X}_L = \mathbf{X} * \mathbf{W}_{\text{conv3x3}} * \mathbf{W}_{\text{pointwise}}.
\end{equation}
This setup captures local relationships while preparing the data for global context modeling.

To incorporate global information, $\mathbf{X}_L$ is divided into $N$ non-overlapping patches $\mathbf{X}_U \in \mathbb{R}^{P \times N \times d}$, where $P = wh$ is the number of pixels in each patch. For each patch $p$, a transformer layer models inter-patch dependencies:
\begin{equation}
    \mathbf{X}_G(p) = \text{Transformer}(\mathbf{X}_U(p)), \quad 1 \leq p \leq P.
\end{equation}
This operation preserves spatial order within and across patches, allowing each pixel to encode both local and global information. After processing, $\mathbf{X}_G$ is folded back into $\mathbf{X}_F \in \mathbb{R}^{H \times W \times d}$ and fused with the original input via a point-wise convolution. This design ensures that the effective receptive field spans the entire spectrogram while maintaining computational efficiency.

\subsection{Lipschitz-Aware Transformer Blocks}
A key innovation of FAST lies in its Lipschitz-aware  blocks, designed to enhance stability by maintaining a bounded Lipschitz constant across the network~\cite{lipsformer}. This is achieved through three main modifications: CenterNorm, Scaled Cosine Similarity Attention (SCSA), and Weighted Residual Shortcuts (WRS).

\subsubsection{CenterNorm}
To ensure Lipschitz continuity and prevent gradient explosion during training, FAST replaces LayerNorm with CenterNorm. Unlike LayerNorm, which normalizes feature vectors by their variance, CenterNorm only subtracts the mean, ensuring stability by keeping the transformation Lipschitz continuous:
\begin{equation}
    \operatorname{CN}(\mathbf{x}) = \boldsymbol{\gamma} \odot \frac{D}{D-1} \left( \boldsymbol{I} - \frac{1}{D} \mathbf{1} \mathbf{1}^{\top} \right) \mathbf{x} + \boldsymbol{\beta},
\end{equation}
where $\mathbf{x}$ represents the input, $\boldsymbol{\gamma}$ and $\boldsymbol{\beta}$ are learnable parameters, and $D$ is the dimensionality of $\mathbf{x}$. This modification stabilizes the gradients without sacrificing model performance.

\subsubsection{Scaled Cosine Similarity Attention (SCSA)}
FAST replaces the conventional dot-product attention mechanism with Scaled Cosine Similarity Attention to maintain stability during training. The attention mechanism is computed as:
\begin{equation}
    \operatorname{SCSA}(\mathbf{Q}, \mathbf{K}, \mathbf{V}) = \nu \mathbf{P} \mathbf{V}, \quad \mathbf{P} = \operatorname{softmax} \left( \tau \mathbf{Q} \mathbf{K}^{\top} \right),
\end{equation}
where $\mathbf{Q}$, $\mathbf{K}$, and $\mathbf{V}$ represent the query, key, and value matrices, and $\nu$ and $\tau$ are predefined or learnable scaling factors. To ensure Lipschitz continuity, $\mathbf{Q}$, $\mathbf{K}$  and $\mathbf{V}$ are normalized as:
\begin{align}
    \mathbf{q}_i &= \frac{\mathbf{x}_i^{\top} \mathbf{W}^Q}{\sqrt{\|\mathbf{x}_i^{\top} \mathbf{W}^Q\|^2 + \epsilon}}, \\
    \mathbf{k}_i &= \frac{\mathbf{x}_i^{\top} \mathbf{W}^K}{\sqrt{\|\mathbf{x}_i^{\top} \mathbf{W}^K\|^2 + \epsilon}}, \\
    \mathbf{v}_i &= \frac{\mathbf{x}_i^{\top} \mathbf{W}^V}{\sqrt{\|\mathbf{x}_i^{\top} \mathbf{W}^V\|^2 + \epsilon}}.
\end{align}

where $\epsilon$ is a smoothing factor to avoid division by zero. This normalization prevents the attention scores from growing uncontrollably, ensuring stable training dynamics.

\subsubsection{Weighted Residual Shortcuts (WRS)}
FAST also modifies the traditional residual connections by introducing Weighted Residual Shortcuts which scale the residual transformation by a learnable parameter $\alpha$ before adding it back to the input:
\begin{equation}
    \operatorname{WRS}(\mathbf{x}, \mathbf{W}) = \mathbf{x} + \alpha \odot f(\mathbf{x}, \mathbf{W}),
\end{equation}
where $\alpha$ is initialized with a small value to control the magnitude of the residual and \(f\) is attention. This scaling ensures that the residual connection does not introduce instability by amplifying variations excessively, thus maintaining a bounded Lipschitz constant.

\subsubsection{Mathematical Formulation for Lipschitz Block}
Finally we define the Lipschitz transformer block combining CenterNorm, SCSA, and WRS as follows: 
\begin{equation}
    \mathbf{x_{i+1}} = \operatorname{CenterNorm}\left(\mathbf{x_i} + \operatorname{DropPath}_{p_i} (\alpha_i f(\mathbf{x_i}))\right),
\end{equation}
where $\alpha_i$ scales the transformation, and DropPath introduces stochastic depth for better generalization. Figure \ref{fig:fast} illustrates the Lipschitz block in the model architecture.

\subsection{Overall Architecture}
The trunk of the FAST model alternates between convolutional layers (specifically, MobileNetV2 blocks) and Lipschitz-aware  blocks. Convolutional layers continue to reduce the spatial dimensions while capturing local features, whereas the Lipschitz transformer blocks integrate global contextual information from the spectrogram.

At the final stage, the model applies global average pooling to aggregate the processed features into a fixed-size vector. This vector is then passed through a series of fully connected layers before being fed into the classification layer. Global average pooling ensures that the most salient features are captured for accurate classification across audio tasks.

The overall architecture of FAST, as depicted in Figure \ref{fig:fast}, demonstrates this integration of convolutional feature extraction with Lipschitz-stabilized  processing, optimized for audio spectrogram analysis.

\section{Experiments}
\label{sec:exp}

\subsection{ADIMA Dataset}

\subsubsection{Dataset Description}
The ADIMA dataset contains 11,775 audio samples (65 hours) in 10 Indic languages (Hindi, Bengali, Punjabi, Haryanvi, Kannada, Odia, Bhojpuri, Gujarati, Tamil, and Malayalam) spoken by 6,446 unique chatroom users. The accompanying task is one of classifying audio samples as ``abusive'' or ``not abusive''. The dataset is well balanced between the two classes - 43.48\% of samples are ``abusive''. The samples of each language are split randomly 70-30 to form the training and test sets. For our experiments on ADIMA, we adhere to the same training pipeline as described in \cite{adima}, ensuring consistency with prior work.

\subsubsection{Monolingual Training}

\begin{table}
\caption{Performance comparison of XLSR-53, AST, HTS-AT, and FAST across different languages (Accuracy/F1 Score). Bold indicates where FAST surpasses other models.}
\small\centering
\begin{tabular}{|c|c|c|c|c|}
\hline
\textbf{Language} & \textbf{XLSR-53} & \textbf{AST} & \textbf{HTS-AT} & \textbf{FAST} \\ 
& \textbf{(Acc/F1)} & \textbf{(Acc/F1)} & \textbf{(Acc/F1)} & \textbf{(Acc/F1)} \\ \hline
Hindi     & 0.77, 0.77 & 0.76, 0.75 & 0.77, 0.76 & \textbf{0.79, 0.79} \\ \hline
Bengali   & 0.81, 0.79 & 0.77, 0.80 & 0.76, 0.80 & 0.78, \textbf{0.83} \\ \hline
Punjabi   & 0.80, 0.82 & 0.85, 0.85 & 0.83, 0.83 & 0.83, 0.83 \\ \hline
Haryanvi  & 0.82, 0.79 & 0.81, 0.80 & 0.82, 0.81 & 0.80, 0.77 \\ \hline
Kannada   & 0.83, 0.79 & 0.81, 0.79 & 0.79, 0.78 & \textbf{0.83, 0.79} \\ \hline
Odia      & 0.83, 0.82 & 0.80, 0.79 & 0.80, 0.79 & 0.81, 0.77 \\ \hline
Bhojpuri  & 0.76, 0.71 & 0.79, 0.80 & 0.80, 0.81 & 0.77, \textbf{0.83} \\ \hline
Gujarati  & 0.79, 0.69 & 0.82, 0.72 & 0.83, 0.73 & \textbf{0.83, 0.73} \\ \hline
Tamil     & 0.80, 0.75 & 0.82, 0.79 & 0.81, 0.79 & 0.80, 0.77 \\ \hline
Malayalam & 0.81, 0.75 & 0.79, 0.76 & 0.79, 0.78 & \textbf{0.82, 0.78} \\ \hline
\end{tabular}
\label{tab:xlsr}
\end{table}

We evaluate the audio classification performance of FAST against the AST, HTS-AT and XLSR-53 models. Table \ref{tab:xlsr} presents comparative results of FAST against other models across various languages in terms of accuracy and F1-scores.

\begin{table}[h]
\caption{Comparison of Model Efficiency. Fewer parameters and quicker inference time together indicate better efficiency.}
\centering
\begin{tabular}{|c|c|c|}
\hline
\textbf{Model} & \textbf{Parameters} & \textbf{Inference Time (secs)} \\ \hline
XLSR-53 & 300M & 0.0579 \\ \hline
AST & 87M & 0.0967 \\ \hline
HTS-AT & 31M & 0.0231 \\ \hline
FAST (Ours) & \textbf{2M} & \textbf{0.0194} \\ \hline
\end{tabular}
\label{tab:model-parameters}
\end{table}

Table \ref{tab:model-parameters} provides a comparative overview of the parameter counts and inference latencies for leading audio classification models. Number of parameters is a crucial metric reflecting the complexity and computational requirements of the model architecture. Inference time is key to real-time usability and mobile device energy consumption, and smaller models do not necessarily result in faster inference. For inference time testing, we follow the single stream scenario discussed in MLPerf Inference Benchmark~\cite{reddi2020mlperfinferencebenchmark}.

We find that FAST not only competes with state-of-the-art models like XLSR-53, AST, and HTS-AT but also surpasses them in several key cases, despite having a more streamlined design with faster inference and two orders of magnitude fewer parameters. Specifically, FAST outperforms or matches these models in both accuracy and F1-score for 4 out of the 10 languages tested—Hindi, Kannada, Gujarati, and Malayalam. In Bengali and Bhojpuri, while XLSR-53 and HTS-AT achieve higher accuracy, FAST leads in F1-score, indicating a more balanced performance across classes. For the remaining languages, FAST holds its ground, demonstrating competitiveness across a diverse range of tasks.

These results demonstrate FAST’s potential in deploying effective audio classification solutions at a fraction of the standard complexity, which may be game-changing for resource-constrained environments such as ADIMA's intended mobile or edge devices. The model's performance across various languages establishes it as a robust option for real-world applications.

\begin{figure}
    \centering
    \includegraphics[width=8.3cm]{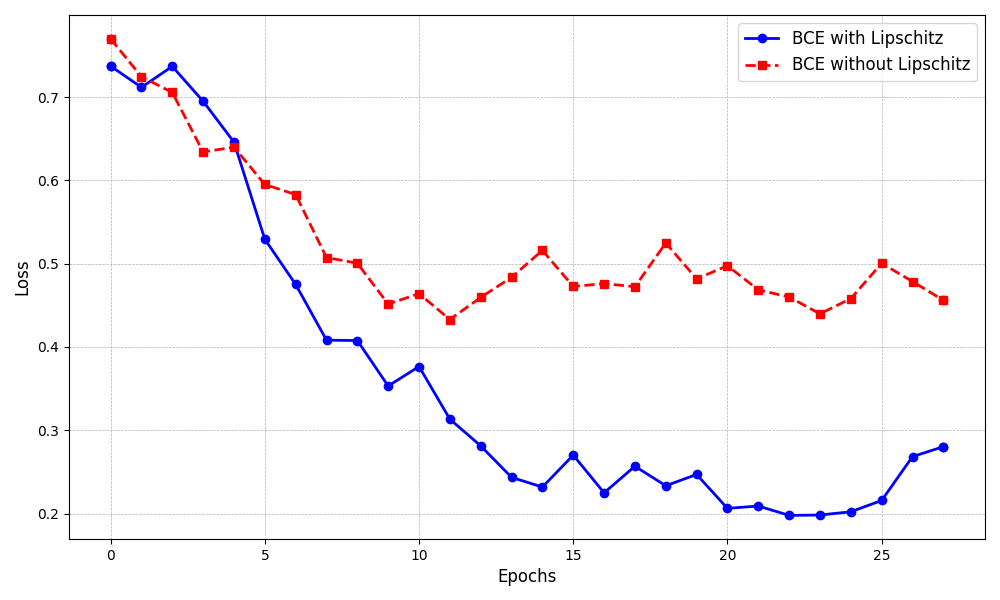}
    \caption{Comparison of training stability and efficiency in FAST with and without Lipschitz continuity components on the ADIMA Hindi language set. Loss is measured with binary cross-entropy (BCE).}
    \label{fig:train-lips}
\end{figure}

\subsubsection{Effect of Lipschitz Components}

We also observe that the integration of Lipschitz continuity components into the FAST architecture benefits training stability and efficiency. During experiments, we found that inclusion of Lipschitz constraints mitigated instability issues in our deep  model. Moreover, training pace was accelerated compared to the version of our model without this component --- Lipschitz regularization appeared to bring quicker convergence, reducing computational overhead and resource requirements. Incorporation of Lipschitz continuity components within FAST not only bolstered its robustness but also streamlined its training dynamics. Figure \ref{fig:train-lips} illustrates the impact of integrating Lipschitz components on training in the FAST architecture for Hindi language. The plot is similar across all languages.

\subsection{AudioSet}

AudioSet \cite{audioset} is a large-scale dataset comprising over 2 million 10-second audio clips sourced from YouTube, each annotated with one or more of 527 possible sound labels. The dataset is organized into three sets: a balanced training set of 22k samples, a full training set containing 2 million samples, and an evaluation set with 20k samples. For our experiments on AudioSet, we adhere to the same training pipeline as described in \cite{ast}, ensuring consistency with prior work.

\begin{table}[h]
\caption{Comparison of mAP scores for AST, HTS-AT, and FAST on AudioSet.}
\centering
\begin{tabular}{|c|c|}
\hline
\textbf{Model Architecture} & \textbf{mAP} \\ \hline
AST & 0.459 \\ \hline
HTS-AT & 0.488 \\ \hline
FAST (Ours) & 0.448 \\ \hline
\end{tabular}
\label{tab:audioset}
\end{table}

\begin{table}[h]
\caption{Model Configuration for FAST.}
\small\centering
\begin{tabular}{|c|c|}
\hline
\textbf{Parameter} & \textbf{Value} \\ \hline
Image Size & (128, 1876) \\ \hline
Hidden Dimensions & (96, 128, 144) \\ \hline
Channels & (16, 32, 48, 48, 64, 64, 80, 80, 96, 96, 384) \\ \hline
Number of Classes & N \\ \hline
Expansion Ratio & 4 \\ \hline
Kernel Size & 3 \\ \hline
Patch Size & (2, 2) \\ \hline
Depths & (2, 4, 4) \\ \hline
\end{tabular}
\label{tab:model_config}
\end{table}

To evaluate the performance of our model, we compare FAST with the state-of-the-art transformer-based models AST and HTS-AT. Table \ref{tab:audioset} presents the mean average precision (mAP) scores for each model. While AST and HTS-AT achieve mAP scores of 0.485 and 0.488, respectively, FAST achieves a slightly lower mAP of 0.448. Despite this difference, FAST maintains competitive performance, particularly considering its lightweight architecture. This underscores the potential of FAST for real-time applications where efficiency and reduced model complexity are critical.

\subsection{Input and Hyperparameter Details}
For ADIMA monolingual training, FAST is trained for 30 epochs using an Adam optimizer with a learning rate of 0.001 and a batch size of 16. The primary evaluation metrics are accuracy and F1-score. In experiments on AudioSet, the model is trained for 5 epochs with an initial learning rate of 1e-5, halved after the second epoch, batch size 16, and using mean average precision (mAP) as the primary metric. The model configuration is shared in Table \ref{tab:model_config}.

\section{Conclusion}
\label{sec:conc}

In this paper, we presented FAST, a novel and lightweight architecture that combines convolutional neural networks with a  backbone, augmented by Lipschitz-based enhancements, for audio classification tasks. Our model demonstrates competitive performance on both the ADIMA and AudioSet tasks, achieving results comparable to state-of-the-art models while utilizing up to 150x fewer parameters. Motivated by the ``live chatroom safety'' origins of ADIMA, we envision FAST to be suitable for a real-time, on-device setting due to its small size and low latency.

For future work, we plan to investigate the potential of transfer learning from vision models to further enhance generalizability to the broader realm of audio classification tasks. Additionally, preliminary experiments suggest that FAST holds promise for few-shot learning across languages using meta-learning techniques like MAML, offering new directions for improving cross-lingual audio tasks.

\bibliographystyle{IEEEtran}
\bibliography{IEEEabrv,mybib}
\end{document}